\documentclass[sigplan,10pt]{acmart}
\usepackage[utf8]{inputenc}

\usepackage{xcolor}
\usepackage{listings}
\usepackage{url}
\usepackage{marginnote}
\usepackage{microtype}

\definecolor{keywordblue}{RGB}{25,25,117}
\definecolor{stringred}{RGB}{159, 34, 34}
\definecolor{commentgreen}{RGB}{29, 138, 29}
\definecolor{emphpurple}{RGB}{79,12,58}

\newcommand{\listingsfont}{\footnotesize\ttfamily}
\acmConference[PROBPROG]{International Conference on Probabilistic Programming}{October 04--06, 2018}{Boston, MA, USA}
\acmYear{\vspace{-30pt}\color{white}{ }}
\acmISBN{} 
\acmDOI{} 

\setcopyright{none}

\title[]{Effect Handling for Composable Program Transformations in Edward2}
\subtitle{Extended Abstract}
\author{Dave Moore}
\affiliation{      
    \institution{Google}            
}
\email{davmre@google.com}          

\author{Maria I. Gorinova}
\affiliation{
    \institution{Google, University of Edinburgh}
}
\email{m.gorinova@ed.ac.uk}          

\begin{document}
\maketitle

\section{Introduction}

Algebraic effects and handlers have emerged in the programming languages community as a convenient, modular abstraction for controlling computational effects. They have found several applications including concurrent programming, meta programming, and more recently, probabilistic programming, as part of Pyro's Poutines library \cite{Pyro}. 

We investigate the use of effect handlers as a lightweight abstraction for implementing probabilistic programming languages (PPLs). We interpret the existing design of Edward2 as an accidental implementation of an effect-handling mechanism, and extend that design to support nested, composable transformations. We demonstrate that this enables straightforward implementation of sophisticated model transformations and inference algorithms.

\vspace{-3pt}
\section{Algebraic Effects and Handlers}

An effectful operation is an operation that interacts with some (possibly external) handling code in order to execute. For example, suppose that a process wants to access a file. It sends a \emph{request} to the OS kernel and suspends execution. The kernel checks the request, executes it, and \emph{responds} with the result of the operation. The process then resumes execution.

Algebraic effects and their handlers \cite{AlgebraicEffects2001, AlgebraicEffectHandlers2009} extend this request-response idea to computations within a program. Impure behaviour arises from a set of effectful operations, whose concrete implementation is separately given in the form of \emph{effect handlers}. The programmer chooses how to handle different operations. Consider an example \cite{Pretnar2015}:
\begin{lstlisting}
let abc = (print('a'); print('b'); print('c'))
let reverse = handler {print(s.k) $\mapsto$ k(); print(s)}
with reverse handle abc
\end{lstlisting}
In this program, \lstinline{abc} is a computation that prints out the letters `a', `b' and `c', in this order, using three separate calls to the \emph{operation} \lstinline{print}. The \emph{handler} \lstinline{reverse} reverses the order in which \lstinline{print} operations are executed: it first resumes the continuation \lstinline{k} to the operation, and only then performs the operation itself. The computation \lstinline{with reverse handle abc} is the result of executing \lstinline{abc}, while handling operations with \lstinline{reverse}: a printout of `c', `b' and `a' in this order.

One very useful feature of effect handlers is that they can be nested to combine the way they interpret the computation. 
%
%
In the presence of a handler \lstinline{join}, which joins the effect of \lstinline{print} statements into a single \lstinline{print} statement, we can easily obtain a reversed single printout `cba', by combining \lstinline{with join handle with reverse handle abc}.

%


\vspace{-3pt}
\section{Effects in Probabilistic Programming}

%
The application of effect handlers to probabilistic programming has been previously discussed \cite{Pyro, Scibior2015} but is perhaps not yet widely appreciated. Consider a Beta-Binomial model:
\begin{lstlisting}
let model(n) = 
    let z = sample(beta(1., 1.), 'z')
    let x = sample(binomial(z, n), 'x')
    return x
\end{lstlisting}
%
%
The insight is to treat sampling statements as operations that can be handled by a separately defined handler.\footnote{Although not covered in this paper, handling deterministic operations can also be desirable, e.g.~for local reparameterization gradients \cite{kingma2015variational}.} This enables a range of useful program transformations, including (though not limited to):
%


\noindent
\textbf{Conditioning.} %
The \lstinline{condition} handler takes a mapping from variable names to their observed values, and changes the respective sampling statements to observe statements:
%
%

\begin{lstlisting}
let condition(name, value) = handler{ 
    sample(dist,name; k) $\mapsto$ 
        k(observe(dist, value, name))}
with condition('x', data) handle model(n)
\end{lstlisting}

\noindent
\textbf{Tracing.} %
A tracing handler accumulates the values of all random variables defined by the model, so that \lstinline{with trace handle model(n)} obtains a sample for both \lstinline{z} and \lstinline{x}. Tracing can also be used for program analysis, for example, computing Markov blankets for efficient algorithms.

\noindent
\textbf{Density function derivation.} Inference algorithms such as Metropolis-Hastings require access to the log joint density $\log p(z, \mathbf{x}) = \log \mathrm{Beta}(z \mid 1, 1) + \sum^{N}_{n=1}\log \mathrm{B}(x_i \mid z)$. This may be derived using a trace-like handler that accumulates the conditional log densities for each value.


\noindent
\textbf{Model reparametrization.} %
\emph{Reparametrizing} a probabilistic model means expressing it in terms of different parameters, and specifying a way to recover the original parameters. A common example, implemented in existing systems such as Pyro \cite{Pyro} and Stan \cite{Stan}, is \emph{unconstraining}: transforming constrained variables such as the Beta variable \lstinline{z} to unconstrained space. This can substantially ease inference and is highly desirable for a system to perform automatically.  

Reparametrization can be expressed using effect handlers. Assuming a handler \lstinline{unconstrain} (expanded below), writing \lstinline{with unconstrain handle model(n)} gives the unconstrained Beta-Binomial model. An elegant aspect of this approach is that it immediately generalizes to other reparameterizations such as non-centering or inverse CDF transformations.

\noindent
\textbf{Variational inference.} %
Effect handling can automatically construct a variational family on-the-fly \citep{wingate2013automated}, e.g., a mean-field variational inference handler may handle \lstinline{sample(dist)} by initialising the parameters \lstinline{mu} and \lstinline{sigma}, and transforming the random draw to \lstinline{sample(normal(mu, sigma))}. Separately, a handler may also be used to align a model's latent variables with samples from a variational model.

\subsection{Composing Effect Handlers}
Many handlers become much more useful when composed; for example, when reparameterizing a model, or automatically constructing a variational model, we would typically then want to derive the joint density function of the transformed model. Composing effect handlers makes this straightforward. For example, the unnormalised posterior on \lstinline{z} is:
\begin{lstlisting}
let posterior(z) = 
  with log_joint(z) handle
    with unconstrain handle 
      with condition(x, data) handle model(n)
\end{lstlisting}

More generally, composing effect handlers allows sophisticated program transformations; for example, the unconstraining, variational guide, and log-joint handlers enable an almost trivial implementation of ADVI \cite{kucukelbir2017automatic}.

\vspace{-3pt}
\section{Effect Handling in Edward2}

Edward2 is a lightweight framework for probabilistic programming in TensorFlow \cite{tran2018}. The main abstraction is the \lstinline{RandomVariable} \lstinline{(RV)}: it wraps a probability distribution with a \lstinline{value} tensor which reifies a sample from that distribution; constructing an \lstinline{RV} implicitly performs a \lstinline{sample} operation. Models are typically
written as generative processes defining a joint distribution over the values of all random variables: 
\begin{lstlisting}
from tensorflow_probability import edward2 as ed
def model(n):
  z = ed.Beta(1., 1., name='z')
  x = ed.Binomial(z, n, name='x')
  return x
\end{lstlisting}
Edward2 supports program transformations by {\em interception}. Running the model in the context of an \lstinline{interceptor} overrides the construction of random variables. To implement this, \lstinline{RV} constructors are wrapped
with a method that checks for an interceptor on a global context stack and, if present, dispatches control.
\begin{lstlisting}
def condition_interceptor(**values):
  def interceptor(rv_constructor, **rv_kwargs):
    rv_name = rv_kwargs['name']
    rv_kwargs['value'] = values.get(rv_name)
    return rv_constructor(**rv_kwargs)
  return interceptor
  
with ed.interception(condition_interceptor(z=0.3)):
  x = model(n)
\end{lstlisting}

We observe that interceptors are essentially an accidental implementation of effect handlers; more specifically, a restricted form in which the handler accesses its continuation only implicitly. The handler (interceptor) overrides a \lstinline{sample} operation with arbitrary computations, potentially including side effects, and ends by invoking an implicit continuation to return a value to the original callsite. Edward2's original framework did not \emph{compose} interceptors, but viewing interceptors as effect handlers suggests that composing them can enable sophisticated program transformations. 

The semantics of composing interceptors may be understood in terms of effect forwarding. Interceptors may call the \lstinline{rv_constructor} directly, in which case the operation is not visible to any higher-level interceptors, or they may explicitly forward the operation by re-wrapping the constructor as \lstinline{interceptable(rv_constructor)}. They may also invoke other \lstinline{RV} constructors, which by default creates wrapped (forwarded) operations.

As an example application requiring nested interceptors, we implement an unconstraining interceptor, using the Bijectors library to handle Jacobian corrections \cite{dillon2017tensorflow}: 
\begin{lstlisting}
from tensorflow_probability import bijectors as tfb
def unconstrain(rv_constructor, **rv_kwargs):
  base_rv = rv_constructor(**rv_kwargs)
  bijector = constraining_transform(base_rv)
  unconstrained_rv = ed.TransformedDistribution(
    distribution=base_rv.distribution,
    bijector=tfb.Invert(bijector))
  return bijector.forward(unconstrained_rv)
\end{lstlisting}
Here the \lstinline{ed.TransformedDistribution} constructor invokes an interceptable operation, while the original \lstinline{base_rv} constructor is not forwarded, so that the transformed program appears to a higher-level handlers (e.g., a log joint density) as containing variables in unconstrained space.

\vspace{-3pt}
\section{Discussion}

%
There is often a gap between theoretical discussions of probabilistic programming and the implementation of practical systems.
We believe the emergence of effect handlers as a convergent design pattern in deep PPLs is notable, and hope that highlighting it may lead to interesting connections in both directions between theory and practice. Compared to Pyro's Poutines \cite{Pyro}, which also implement effect handling, Edward2's interception provides substantially similar functionality with a different, somewhat lighter-weight interface. We believe both are interesting points in design space and look forward to exploring their tradeoffs.

We do not claim that effect handling is a complete mechanism for probabilistic programming; for example, it is not obvious how non-local rewrites such as general symbolic algebra on computation graphs \cite{autoconj2018} might fit in an effect-handling framework. Understanding the space of program transformations that can be usefully specified as effect handlers is an exciting area of future work.

\bibliography{bib}

\end{document}